\documentclass{aa}

\usepackage{graphicx,amsmath}
\usepackage{units}
\usepackage{natbib}
\usepackage{txfonts}
\usepackage[usenames,dvipsnames]{color}

\newcommand{\micron}{\mu m}
\newcommand{\msun}{M_\odot}
\newcommand{\dd}[1]{{\mathrm{d}#1}}
\newcommand{\sub}[2]{#1_\mathrm{#2}}
\newcommand{\nnh}{$\mathrm{N_2H^+}$}
\renewcommand{\deg}{^{\circ}}

\defcitealias{che12}{CAD12}

\authorrunning{Schmalzl et.\ al (2013)}
\titlerunning{The structure of the star-forming core CB\,17}

\begin{document}

\title{The Earliest Phases of Star formation (EPoS):}
\subtitle{Temperature, density,
 and kinematic structure of the star-forming core CB\,17\thanks{Data of Figs.~\ref{f-02} and \ref{f-03}
are available in electronic form at the CDS via anonymous ftp to cdsarc.u-strasbg.fr (130.79.128.5)
or via {\tt http://cdsweb.u-strasbg.fr/cgi-bin/qcat?J/A+A/}}}

\author{M.\ Schmalzl\inst{1,2,3}
        \and
        R.\ Launhardt\inst{2}
        \and
        A.M.\ Stutz\inst{2}
        \and
        H. Linz\inst{2}
        \and
        T.L.\ Bourke\inst{3}
        \and
        H.\ Beuther\inst{2}
        \and
        Th.\ Henning\inst{2}
        \and
        O.\ Krause\inst{2}
        \and
        M.\ Nielbock\inst{2}
        \and
        A.\ Schmiedeke\inst{2,4}
}

\institute{
        Leiden Observatory, Leiden University, P.O.\ Box 9513, 2300 RA, Leiden, The Netherlands\\
        \email{schmalzl@strw.leidenuniv.nl}
        \and
        Max Planck Institute for Astronomy, K\"onigstuhl 17, 69117 Heidelberg, Germany
        \and
        Harvard-Smithsonian Center for Astrophysics, 60 Garden Street, Cambridge, MA 02138, USA
        \and
        Universit\"at zu K\"oln, Z\"ulpicher Stra{\ss}e 77, 50937 K\"oln, Germany
}

\date{Received 29 June 2013 / Accepted 25 May 2014}

\abstract 
{The initial conditions for the gravitational collapse of molecular
cloud cores and the subsequent birth of stars are still not
well constrained. The characteristic cold temperatures ($\sim\unit[10]{K}$) in
such regions require observations at sub-millimetre and longer
wavelengths. The \protect{\it Herschel Space Observatory} and
complementary ground-based observations presented in this paper have
the unprecedented potential to reveal the structure and kinematics of
a prototypical core region at the onset of stellar birth.
}
{This paper aims to determine the density, temperature, and velocity
structure of the star-forming Bok globule CB\,17. This isolated region
is known to host (at least) two sources at different evolutionary
stages: a dense core, SMM1, and a Class~I protostar, IRS.
}
{We modeled the cold dust emission maps from $\unit[100]{\micron}$
to $\unit[1.2]{mm}$ with both a modified blackbody technique to
determine the optical depth-weighted
line-of-sight temperature and column density and a ray-tracing technique to
determine the core temperature and volume density structure.
Furthermore, we analysed the
kinematics of CB17 using the high-density gas tracer \nnh{}.
}
{From the ray-tracing analysis, we find a temperature in the
centre of SMM1 of $T_0 = \unit[10.6]{K}$, a
flat density profile with radius $\unit[9.5\times10^3]{au}$, and a central
volume density of $\sub{n}{H,0}=\unit[2.3\times10^5]{cm^{-3}}$.
The velocity structure of the \nnh{} observations
reveal global rotation with a velocity gradient of $\unit[4.3]{km\,s^{-1}\,pc^{-1}}$.
Superposed on this rotation signature we find a more complex velocity field,
which may be indicative of differential motions within the dense core.
}
{SMM is a core in an early evolutionary stage at the verge of being bound,
but the question of whether it is a starless or a protostellar core remains
unanswered.}


\maketitle

\section{Introduction}

Knowledge of the initial conditions
is crucial for understanding the evolution from starless to prestellar to
protostellar cores \citep[e.g.,][]{dif07}. Owing
to their low temperatures of $T\sim\unit[10]{K}$ \citep[e.g.,][]{ber07,and13},
the peak of the thermal dust emission of prestellar cores is 
found at sub-millimetre wavelengths.
Observations with millimetre and sub-millimetre bolometer
arrays revealed a flattening of the radial intensity distribution in
the central regions \citep[e.g.,][]{war94,war99,shi00}, which
contradicted the assumption of a singular isothermal sphere
\citep[e.g.,][]{shu77} as the initial condition of protostellar
collapse. The degeneracy of temperature and column density that
cannot be resolved with observations from the Rayleigh-Jeans regime only meant that
this flattening was
(under the assumption of isothermal spheres) naturally interpreted
to originate in the density profiles.

External heating
by the interstellar radiation field (ISRF) will, however, result in temperature
profiles that decline toward the centre as the level
of shielding by dust grains increases \citep[e.g.,][]{leu75}.
\citet{eva01} used radiative transfer calculations with ISRF heating to
simulate SEDs for both isothermal cores and
cores with temperature gradients. They find that the observed SEDs
can also be explained with non-isothermal cores
\citep[see also][]{zuc01}, which naturally results in
smaller flat regions and enhanced central volume densities
as compared to isothermal cores. In flux
ratio maps of 170 and $\unit[200]{\micron}$, a wavelength range well
outside the Rayleigh-Jeans regime for temperatures of the order of $\unit[10]{K}$,
\citet{war02} found observational evidence for such
temperature gradients towards the centre of starless cores.

Knowledge of the temperature distribution throughout the
molecular cloud cores is essential when it comes to understanding a few
key processes in star formation, such as the appearance of molecules as a
result of grain-surface chemistry \citep[e.g.,][]{taq13} or the
excitation and radiation of gas-phase molecules \citep{pav07}.
Additionally, the thermal energy
is the major contributor to the stability of a core against
gravitational collapse \citep[e.g.,][]{lau13}. Low temperatures can promote
gravitational instability in cores and inevitably lead -- in the absence
of other stabilising factors - to their collapse. During the
early stage, these prestellar cores do show infall motions, but appear starless
\citep[e.g.,][]{cra05,cas12}. Once the central region with its enhanced density
becomes opaque to infrared radiation, cooling becomes less efficient. Further
compression is accompanied by a temperature rise, which leads to the formation
of a so-called first hydrostatic core
\citep[FHSC; e.g.,][]{lar69,mas08,tom10,com12} in a core, which is now of
protostellar nature.
Owing to their deeply embedded nature, low intrinsic luminosity, and short
lifetime, FHSCs are hard to detect and still have to be unambiguously and
observationally identified. Indeed,
distinguishing these early phases of protostellar evolution has
proven observationally challenging. Some FHSC candidates have been found
so far \citep{bel06,eno10, che10, pin11, che12, pez12}, but none of these could be
confirmed yet. Additional confusion for an unambiguous detection is caused
by the fact that simulations of FHSC formation lead -- depending on
initial conditions and underlying physical processes -- to different
FHSC lifetimes, sizes, and luminosities, and this eventually leads
to ambiguous predictions for observations
\citep[e.g.,][]{omu07,mas08,sai08,tom10,com12}. 
As \citet{eno10} point
out, the observational signature of a VeLLO (Very Low Luminosity Object, e.g.,
\citealt{you04}) could originate 
in either an FHSC or a highly embedded faint Class 0 protostar in a quiescent
accretion phase \citep[e.g.,][]{stu13}.

The launch of the \textit{Herschel Space Observatory}\footnote{Herschel is
an ESA space observatory with science
instruments provided by European-led Principal Investigator
consortia and with important participation from NASA.}
\citep{pil10} with its photometer wavelength coverage from 70 to
$\unit[500]{\micron}$ allowed astronomers to view the
earliest stages of
star formation with unprecedented sensitivity and spatial resolution
\citep[e.g.,][]{mol10,and10,stu10,arz11,stu13}. 
The \textit{Herschel} Key Programme EPoS (Earliest Phases of
Star formation, PI: O.\ Krause) is dedicated to observations of both
high-mass star-forming regions
\citep[e.g.,][]{beu10,beu13,hen10,lin10,rag12,pit13,kai13} and
isolated low-mass pre- and protostellar cores
\citep[e.g.,][]{stu10,nie12,lau13,lip13}. The focus for the low-mass
part of the project is mainly on deriving the
density and dust
temperature structure in the early phases
of protostellar evolution and on determining the initial
conditions of protostellar collapse. Although most stars form in
clustered environments \citep{lad03,eva09}, the
EPoS globule sample specifically focusses on nearby, isolated cores because
they offer a pristine, undisturbed view of the basic processes that
lead to the formation of protostars in the absence of complicating
factors, such as competitive accretion, tidal interactions, and
stellar feedback.

\begin{figure}[tb]
 \includegraphics{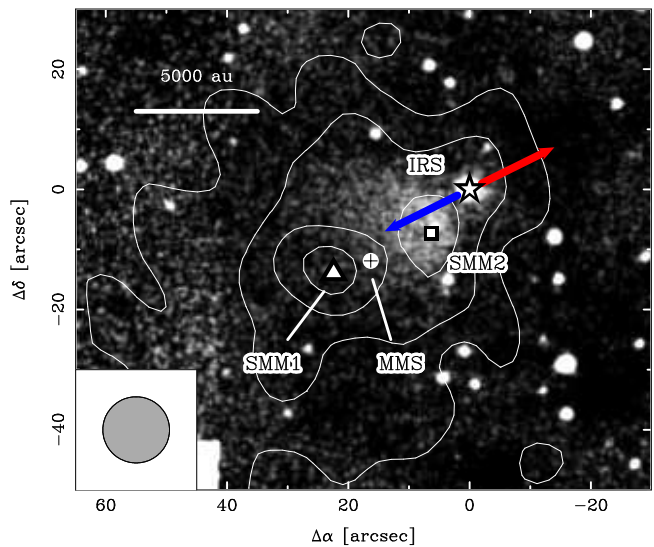}
 \caption{\label{f-01} $\unit[1.2]{mm}$ dust continuum emission
 (contours, plotted in steps of $\unit[3]{\sigma}$) overplotted on a
 NIR $K$ image of CB\,17 \citep{lau10}. The arrows represent the red- and
 blue-shifted components of the outflow detected by \citet{che12}.
 The grey circle in the bottom left corner represents the angular
 resolution of the $\unit[1.2]{mm}$ dust continuum image.
 The coordinate offsets are given w.r.t.\ to the
 position of IRS
 ($\alpha_\mathrm{J2000}=04^\mathrm{h}04^\mathrm{m}33^\mathrm{s}\!\!.76$,
 $\delta_\mathrm{J2000}=+56^\circ56^\prime16^{\prime\prime}\!\!.5$,
 \citealt{che12}).}
\end{figure}

\begin{figure*}[tb]
 \includegraphics{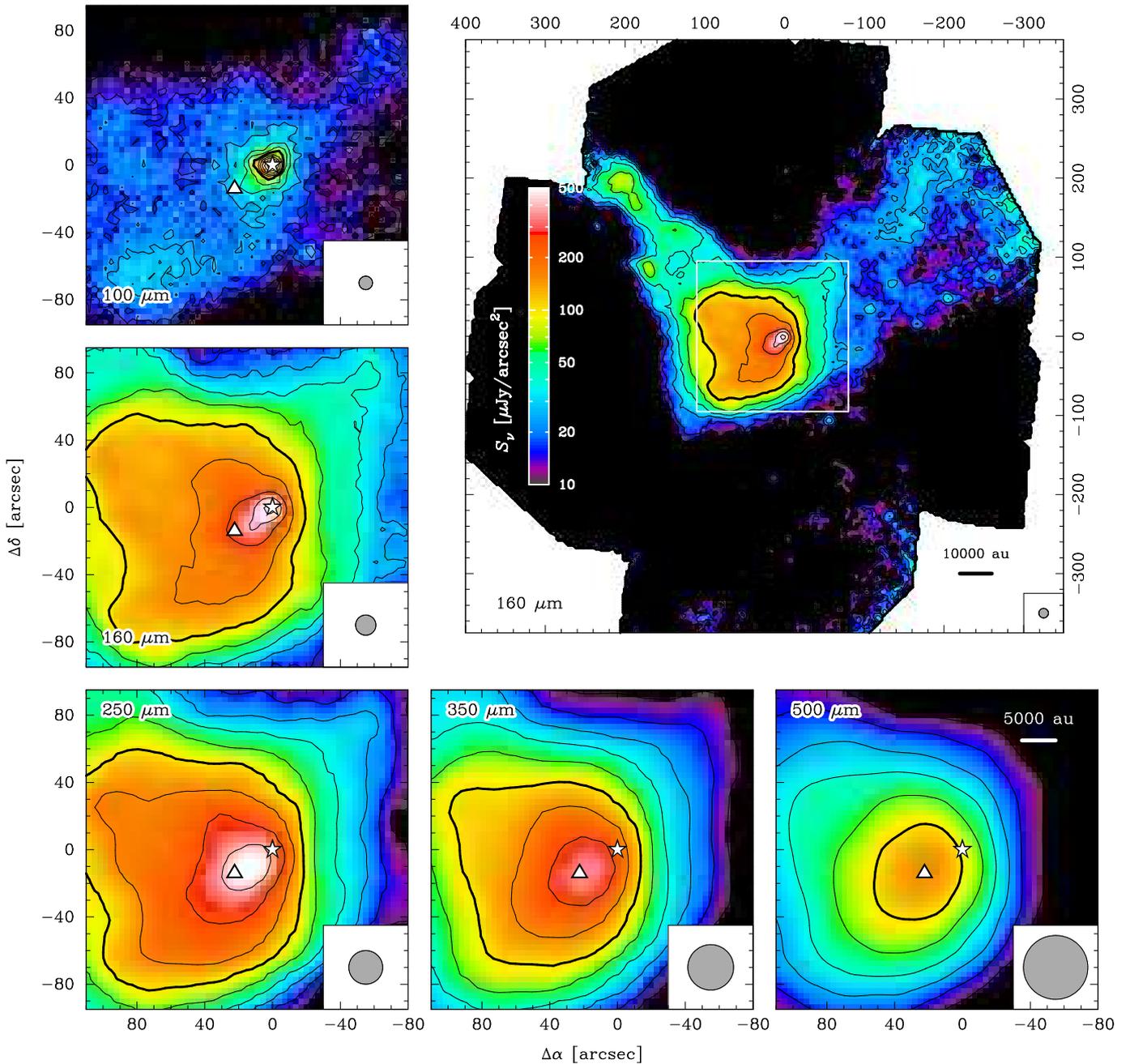}
 \caption{\label{f-02} \textit{Herschel} dust emission
  maps at $\unit[100]{\micron}$, $\unit[160]{\micron}$,
  $\unit[250]{\micron}$, $\unit[350]{\micron}$, and $\unit[500]{\mu
  m}$. The big panel represents the full field of view at
  $\unit[160]{\micron}$, which shows the
  cometary structure of CB\,17. The small panels show detailed views of
  the white rectangular area for all \textit{Herschel}
  observations, ranging from $\unit[100]{\micron}$ to
  $\unit[500]{\micron}$. The surface brightness scale is the same for
  all images. The contours range from $\unit[10]{\mu
  Jy/arcsec^2}$ up in logarithmic steps of
  $\unit[0.2]{dex}$. The thick contour represents
  $\unit[100]{\mu Jy/arcsec^2}$. The respective beam sizes are
  depicted in the bottom right corners. Symbols mark the
  positions of SMM1 ($\blacktriangle$) and IRS
  ($\bigstar$).}
\end{figure*}

In this paper we present an observational study towards
the EPoS target CB\,17 \citep[LDN\,1389,][]{cle88, lyn62}.
This nearby, isolated Bok globule is located at an estimated distance
of $\unit[250\pm50]{pc}$ and hosts a Class I protostar (IRS, accompanied by
a NIR nebula), and an adjacent dense core (SMM) at an angular separation of
$\sim\unit[30]{\arcsec}$ or $\unit[5000]{au}$ \citep{lau10}.
In sensitive 1.2-mm single-dish observations the emission around SMM
exhibits a main peak (SMM1), but also shows
a considerable extension towards the north-west
(SMM2, see Figure~\ref{f-01}). Thanks to detections
in all \textit{Spitzer} MIPS bands and ground-based NIR observations, which
allow a determination of the bolometric temperature \citep{lau10}, and
the detection of low-velocity outflows \citep{che12}, the classification of IRS is well established
by now. In contrast, the evolutionary stage of
SMM is still under debate. Detections of N$_2$H$^+$
\citep{cas02a} and NH$_3$ \citep{lem96}, and a prominent $\unit[8]{\mu m}$-shadow
at the position of strongest emission at mm-wavelengths \citep{lau10} indicate
that SMM has a starless nature. Recent studies by \citet{lau13}
and \citet{lip13} suggest that it is marginally bound and thus a prestellar core.
In interferometric observations, \citet{che12} detected a low S/N, compact
$\unit[1.3]{mm}$ continuum source, MMS, within the boundaries of
SMM (Figure~\ref{f-01}). Thanks to this
detection in concert with a possible detection of a low-velocity outflow,
\citet{che12} classified MMS as a candidate FHSC, which would then mean that
SMM is instead a protostellar core.

The goal of this paper is to perform an in-depth analysis of CB\,17 and
to reveal the evolutionary state of SMM. For this purpose we used
FIR \textit{Herschel} dust emission maps \citep{lau10,lip13} and tailored our
analysis to accommodate the morphological characteristics of CB\,17. Moreover,
we used interferometric observations of the cold gas-tracer N$_2$H$^+$ to
spatially resolve the kinematic structure of the dense core.

This paper is
structured as follows. In Section~\ref{s-obs} we describe our
observations and data. The temperature and density structure of CB\,17, which
is derived from the continuum data, is presented in
Section~\ref{s-profiles}. In Section~\ref{s-res} we focus on
the small-scale structure and kinematics
within the dense core. The findings from the two previous sections are then
discussed in Section~\ref{s-discussion}, which is followed by
a summary in Section~\ref{s-summary}.

\section{Observations}
\label{s-obs}

\subsection{Dust continuum emission}

\subsubsection{\textit{Herschel}}

CB\,17 was observed with SPIRE \citep{gri10} at 250, 350,~and $\unit[500]{\micron}$
and with PACS \citep{pog10} at 100~and $\unit[160]{\micron}$ as
part of the \textit{Herschel} Guaranteed Time Key Programme EPoS on
February 13 and 23, 2010, respectively. Observations with PACS were
obtained with two scan directions oriented perpendicular to each other
to eliminate striping in the final combined maps of effective size
$\sim(\unit[11]{\arcmin})^2$. The FWHM beam sizes are $\unit[7.1]{\arcsec}$
at $\unit[100]{\micron}$ and $\unit[11.2]{\arcsec}$ at $\unit[160]{\mu
 m}$ \citep{ani11}. Likewise, the SPIRE observations were obtained as
a scan map, covering an area of $\sim(\unit[18]{\arcmin})^2$. The
approximate FWHM beam sizes are $18.2$, $25.0$, and
$\unit[36.4]{\arcsec}$ at 250, 350, and $\unit[500]{\micron}$,
respectively \citep{ani11}. A detailed description of these observations
and the respective data reduction
for all targets within the EPoS sample can be found in
\citet{lau13}.

\subsubsection{Additional sub-mm observations}

In addition to the \textit{Herschel} observations, which cover the
wavelength range from 100 to $\unit[500]{\micron}$, we used SCUBA
observations at $\unit[850]{\micron}$ (Programme~ID: M98BC21, beam size:
$\unit[14.3]{\arcsec}$) and IRAM~30m continuum observations at
$\unit[1.2]{mm}$ (beam size: $\unit[10.9]{\arcsec}$). A detailed
overview of these observations can be found in \citet{lau10} and \citet{lau13}.

\subsection{Molecular line emission}

Complementary to the observations of thermal dust emission with
\textit{Herschel}, we also observed line radiation from \nnh. This molecule is known to be 
a tracer of the structure and kinematics
of dense cores \citep{joh10}. This is mainly because
CO as a C-bearing molecule 
that can destroy \nnh  \ freezes out onto the dust grains in the cold
dense cores, whereas N-bearing molecules are less prone to this
effect. Furthermore, with its critical density of
$\sub{n}{H,crit}\sim\unit[10^5]{cm^{-3}}$,
\nnh(1--0) emits most strongly at the densities that are reached in
low-mass prestellar cores \citep{lau10}.

Observations of the \nnh(1--0) hyperfine structure (HFS)
complex at $\unit[93.171]{GHz}$ were obtained with the \textit{Plateau de
 Bure Interferometer} (PdBI) on 2 and 8 April 2009 in configurations
6Cq and 6Dq, respectively. 3C84 served as bandpass and flux density
calibrator, and 0444+634 and 0355+508 served as gain
calibrators. The spectrometer was set up to have a bandwidth of
$\unit[20]{MHz}$, with each of its 512 channels having a width of
$\unit[39]{kHz}$. At the frequency of the \nnh(1--0) transition, the
total bandwidth then conforms to a velocity range of
$\unit[64]{km\,s^{-1}}$ and a channel separation of
$\unit[0.13]{km\,s^{-1}}$ (and thus an effective spectral resolution
of $\unit[0.24]{km\,s^{-1}}$). Therefore, we were easily able to
capture the full extent of the HFS complex, which stretches out over
$\unit[15]{km\,s^{-1}}$.

In addition to the interferometric observations, we used single-dish
data to account for the missing short spacings. This data was obtained
with the IRAM-30m antenna with OTF mapping on 12 July 2004. The data
is remapped to match the phase centre of the PdBI observations, and
then merged with the interferometric data.
Calibration, combination of single-dish with interferometric data,
imaging, and image deconvolution are all performed by using
{\tt GILDAS}\footnote{http://www.iram.fr/IRAMFR/GILDAS/} software. 
Natural weighting of the \textit{uv}-data results in a synthesised FWHM beam size of
$\unit[4.5]{\arcsec}\times\unit[3.7]{\arcsec}$ 
and a sensitivity of $\unit[10]{mJy\,beam^{-1}}$ per
channel.

\begin{figure*}[bt]
 \includegraphics{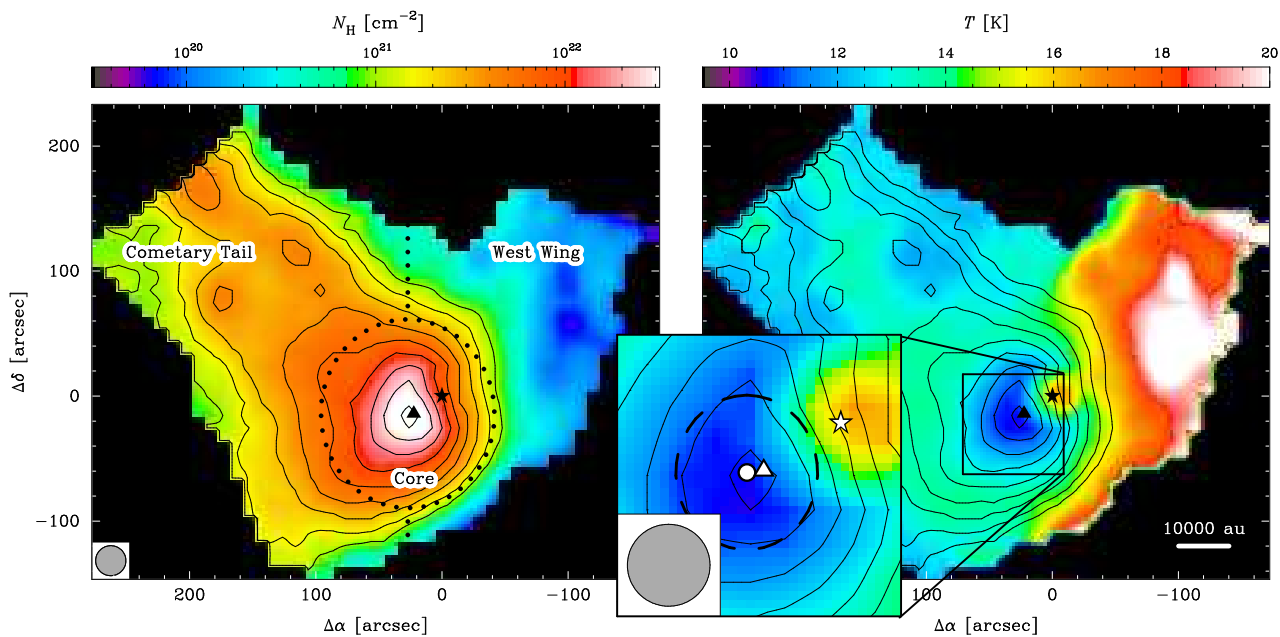}
 \caption{\label{f-03} Hydrogen column density (left
 panel) and dust temperature maps (right panel) of
 CB\,17. The dotted lines mark the boundaries between the three
 regions discussed in the text (core, cometary tail, west wing). The data for the
 \textit{cometary tail} and \textit{west wing} are fitted by using
 the MBB algorithm, whereas
 the \textit{core} is fitted using the RT-procedure (Section~\ref{s-rt}).
 The temperature in the\textit{
 core} represents the mid-plane temperature (i.e., the minimum temperature
 along the line-of-sight), whereas in the
 other two regions the LoS-averaged dust
 temperature is displayed. Symbols mark the
 positions of SMM1 ($\blacktriangle$) and IRS ($\bigstar$) from \citet{lau10}.
 The inset shows a zoom towards the cold and dense regions. The newly derived
 postition of SMM1 and the FWHM of the volume density distribution
 are marked with a white circle
 and dashed line, respectively.
 The contours in all panels represent the column
 density, starting from $\sub{N}{H}=\unit[10^{21}]{cm^{-2}}$,
 and subsequent levels in steps of $\unit[0.2]{dex}$. }
\end{figure*}

\section{Large scale temperature and density distribution}

\label{s-profiles}

\subsection{Overview}

The \textit{Herschel} observations, which are shown in
Figure~\ref{f-02}, give an overview of the thermal
dust emission of CB\,17. The bulk of the emission is found around
SMM and IRS. A long cometary tail extending towards the north-east can
easily be seen
at all wavelengths $>\unit[100]{\micron}$. The $\unit[160]{\micron}$ image
shows a set of bright knots of emission along 
this tail, a feature that is washed out with increasing
wavelength owing to the decreasing spatial resolution.

The detailed views in Figure~\ref{f-02} show that
the Class~I protostar IRS is the dominant source at wavelengths shorter than
$\unit[200]{\micron}$, whereas at longer wavelengths the bulk of the emission
comes from the dense core SMM. This shift in the emission peak with
wavelength has also been observed in CB\,244, which hosts a Class~0 protostar
and a pre-stellar core at an angular separation of $\sim\unit[100]{\arcsec}$
\citep{stu10}. It can be clearly attributed to the different evolutionary
stages. Spectral energy distributions (SED) of protostellar cores are, compared to
starless cores, characterised by higher temperatures, which results in
a shift of the emission peak towards shorter wavelengths.
In CB\,17, this effect is less clear because of the small angular separation between
IRS and SMM, which leads to considerable blending
of the emission from these two sources.

\subsection{Dust temperature and hydrogen volume density profiles}\label{s-rt}

In the pre-\textit{Herschel} era, dust emission was often used to
estimate column densities and masses through single-wavelength observations,
using an educated guess for the
dust temperature and dust opacity \citep[e.g.,][]{lau97,mot98}.
With \textit{Herschel} it is possible to
use both the spectral and spatial information of the dust continuum emission to model
the density and dust temperature distribution.
For our modelling, the available \textit{Herschel} observations
were complemented by SCUBA and IRAM~30m data, which extends the
wavelength coverage from $\unit[100]{\micron}$ to $\unit[1.2]{mm}$. Thus,
we fully sample the peak of the SED.

To determine the dust temperature and hydrogen volume density profiles,
we closely follow the procedures as described in \citet{lau13}
and \citet{lip13}. We use dust opacities from \citet{oss94} for
coagulated dust with thin ice mantles ($\unit[0.1]{Myr}$ of coagulation time
at a gas density of $\unit[10^5]{cm^{-3}}$) and a hydrogen gas-to-dust mass
ratio of 110 \citep{sod97}. In contrast to previous studies by \citet{lau13}
and \citet{lip13}, which also
included CB\,17, we are deliberately omitting the observations at
$\unit[500]{\mu m}$ to gain
a factor of 1.5 in terms of spatial resolution. All our images are then
convolved to the \textit{Herschel} beam of $\unit[350]{\mu m}$
($\unit[25]{\arcsec}$ or $\unit[6\,250]{au}$). This allows us to reduce
the blending of the emission from the dense core and the Class~I source.
Moreover, by using a higher spatial
resolution we aim to minimise convolution effects for a bona-fide determination of
a flattening of the volume density profile towards the centre.

To determine the temperature, column, and volume density structure of CB\,17,
we apply two different techniques. The modified blackbody (MBB) routine gives
us line-of-sight averaged quantities, whereas the ray-tracing (RT) approach
accounts for temperature and
volume density gradients along the line-of-sight.
The radial temperature profile in the RT-model is defined as
\begin{align}
        \label{e-01}
        T(r)=\sub{T}{core}-\Delta T\,\left(1-\mathrm{e}^{-\tau_\mathrm{ISRF}(r)}\right)
\end{align}
where $\Delta T=\sub{T}{core}-\sub{T}{min}$. $\sub{T}{core}$ represents the
temperature at the core edge, and $\sub{T}{min}$ is the minimum temperature.
The term
\begin{align}
        \label{e-02}
        \tau_\mathrm{ISRF}(r)=\tau_\mathrm{ISRF,0}\,\,\frac{\int_r^{r_\mathrm{core}}\sub{n}{H}(r^\prime)\,\dd{r^\prime}}{\int_0^{\sub{r}{core}}\,\sub{n}{H}(r^\prime)\dd{r^\prime}}
\end{align}
accounts for the a priori unknown mean dust opacity and the UV shape of the
ISRF, where $\sub{\tau}{ISRF,0}$ is an empirical scaling parameter.
The temperature in the core centre $T_0$
is then calculated by
\begin{align}
        T_0=\sub{T}{core} - \Delta T\, (1-\mathrm{e}^{-\tau_\mathrm{ISRF,0}}),
\end{align}
which converges towards $\sub{T}{min}$ for $\sub{\tau}{ISRF,0}\gg1$.

In our analysis, the hydrogen volume density profile is defined as
\begin{align}
        \label{e-03}
        \sub{n}{H}(r)=\frac{\sub{n}{H,0}}{\left[1+(r/r_0)^2\right]^{\eta/2}}
\end{align}
where $\sub{n}{H,0}$ is the density of hydrogen nuclei in the core
centre, $r_0$ is the radius of the central density plateau, and $\eta$ the power-law slope
at large radii. We cut off the density profile at a radius of
$\sub{r}{out}=\unit[30\,000]{au}$ at volume densities of a few
$\unit[10^2]{cm^{-3}}$. The value of $\sub{r}{out}$ is unconstrained,
but setting a cutoff at larger radii only has a marginal effect on the
best-fit parameters.

The hydrogen column density and temperature maps are shown in Figure~\ref{f-03}.
Towards the regions of highest density (\textit{Core})
we applied the RT approach. Owing to the apparent
breakdown of spherical symmetry at $r\gtrsim\unit[18,000]{au}$
(which can be seen in the radial profiles discussed later in this paragraph),
the map shows the line-of-sight averaged quantities from the MBB fitting outside
the core.
That no jumps are seen in temperature or column density at the transition
between the core (RT) and envelope (MBB) confirms that temperature gradients are
only important within the core region and that the MBB approximation
is therefore appropriate for the envelope.

The column density peak, which is also accompanied by a temperature minimum,
is obviously offset from IRS. We see no indication of
a local column density enhancement at the position of the Class~I source, which
suggests that it is not deeply embedded, but most likely located in front of
or behind the dense core, and devoid of significant amounts of circumstellar
material. We attribute the offset between the position of IRS
and the temperature peak as being the result
of blending the emission of the cold, dense core with the Class~I source
(see inset in Figure~\ref{f-03}).
This could also compromise determination of the position of
the dense core, which is defined as the column-density weighted mean position.
We estimated a positional uncertainty of the dense core to be of the order
of $\unit[10]{\arcsec}$. Despite this uncertainty, the position of the column
density peak and temperature minimum are spatially coeval with the emission
peak SMM1 in the $\unit[1.2]{mm}$ continuum \citep{lau10}. Henceforth, we
denote the dense core as SMM1.

\begin{figure}[tb]
 \includegraphics{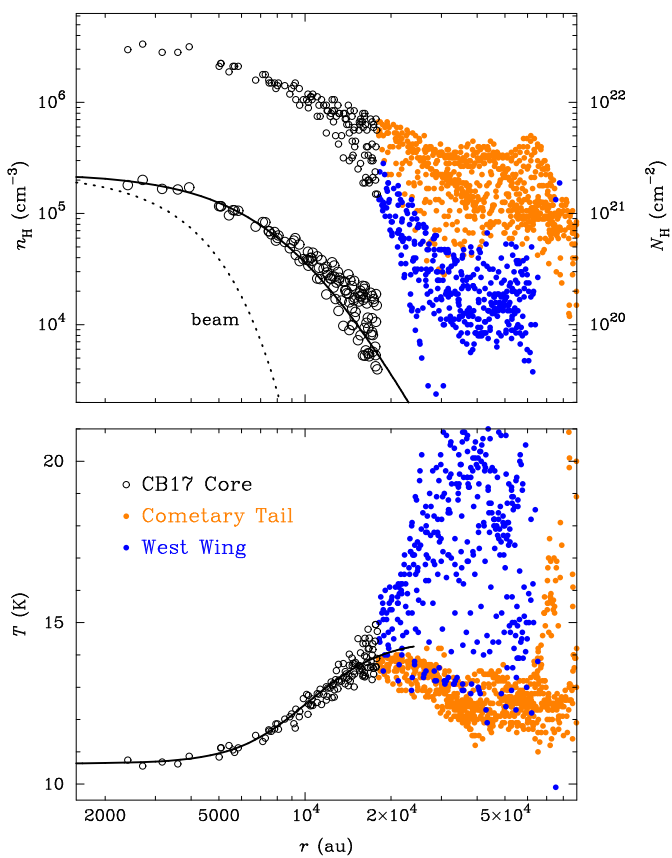}
 \caption{\label{f-04}Azimuthally averaged
 column density (top, small symbols), volume density (top, large symbols),
 and temperature (bottom) distributions. The data for the
 \textit{cometary tail} and the \textit{west wing} are taken from the MBB,
 whereas the data for the \textit{core} is the result of the RT fit
 (Section~\ref{s-rt}). The pixels within a beam size from the Class~I protostar
 IRS are excluded. The solid line shows the curves for
 the best-fit model (Table~\ref{t-01}), the dotted line indicates the
 beam size.}
\end{figure}

The radial dust temperature and hydrogen volume density profiles of SMM1
are shown in
Figure~\ref{f-04}, and the best-fit parameters are listed in
Table~\ref{t-01}. The breakdown of the spherical symmetry
is particularly obvious in these plots. We separate CB\,17 into three
distinct regions: the \textit{core}, the warm \textit{west wing},
and the \textit{cometary tail} (see Figure~\ref{f-03}).

The dense core SMM1 is characterised by a flat inner plateau with
$r_0\sim\unit[9\,500]{au}$ and a power-law index of $\eta=4.9$, which conforms
to a FWHM of the core volume density distribution of
$\unit[10\,900]{au}$. Since this is well beyond the
spatial resolution ($\sim\unit[6\,000]{au}$), we conclude that the flattening
towards the centre is not an artefact of the image convolution. This indicates
an early evolutionary stage, since a flat volume density profile is
characteristic of prestellar cores, whereas protostellar cores exhibit
power-law density distributions down to radii well below our resolution limit
\citep[e.g.][]{jor02}. The central
volume density is derived to be $\sub{n}{H}=\unit[2.3\times10^5]{cm^{-3}}$.
In contrast
to \citet{lip13}, we derive by a factor of $\sim1.5$
higher column and volume densities, and a slightly smaller plateau radius.
These differences can be attributed entirely to the different spatial
resolutions that were used in these two studies.

The derived parameters
and error bars assume the single
dust model that we used in this paper. An analysis and discussion of how
different dust models might affect the derived parameters can be found in
\citet{lau13} and \citet{lip13}.

\begin{table}[bt]
\caption{CB\,17-SMM1 dense core parameters.}
\label{t-01}
\centering
\begin{tabular}{ll}
\hline\hline
\multicolumn{2}{c}{LoS-averaged fit parameters\tablefootmark{$\dagger$}} \\ \hline
$\alpha_0$ (J2000) & $04^\mathrm{h}04^\mathrm{m}37^\mathrm{s}\!\!.1\pm\unit[10]{\arcsec}$ \\
$\delta_0$ (J2000) & $+56^\circ56^\prime02^{\prime\prime}\pm\unit[10]{\arcsec}$ \\
aspect ratio & 1.1 \\
$P.A.$ & 179$^\circ$\\
$\sub{r}{out}$ & $\unit[30\times10^3]{au}$ \\
$\sub{r}{core}$ & $\unit[18\times10^3]{au}$ \\
$\sub{T}{core}$ & $\unit[14.2\pm0.7]{K}$ \\
\hline \hline \multicolumn{2}{c}{ray-tracing fit parameters\tablefootmark{$\ddagger$}} \\ \hline
$r_0$ & $\unit[(9.5\pm0.6)\times10^3]{au}$ \\
$\eta$ & $4.9\pm0.4$ \\
$\sub{\tau}{ISRF,0}$ & $6.5\pm0.7$ \\
$T_0$ & $\unit[10.6\pm0.3]{K}$ \\
$\sub{n}{H,0}$ & $\unit[(2.3\pm0.2)\times 10^5]{cm^{-3}}$ \\
$\sub{N}{H,0}$ & $\unit[(4.3\pm0.2)\times 10^{22}]{cm^{-2}}$ \\
$\sub{M}{core}$ & $\unit[(2.3\pm0.3)]{\msun}$ \\
\hline \\
\end{tabular}
\tablefoot{ 
\tablefoottext{$\dagger$}{These parameters are derived from the column
 density and dust temperature map derived through the LoS-averaged
 fitting, and are kept constant during the ray-tracing
 fitting.}\\
\tablefoottext{$\ddagger$}{These parameters are derived from the
 azimuthally averaged mid-plane volume density and dust temperature
 profiles (Figure~\ref{f-04}) after each iteration
 cycle. The error bars are only representative of the used
dust model. A more general discussion about how different dust
models influence the best-fit parameters can be found in \citet{lip13}.}}
\end{table}

\section{A high-resolution view of SMM1}
\label{s-res}

\subsection{Dust continuum emission}
\label{s-firs}

\begin{figure}[tb]
 \includegraphics{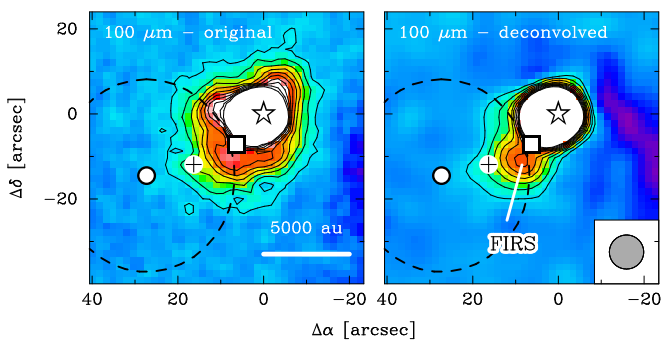}
 \caption{\label{f-05} Original (left) and
 deconvolved (right panel) $\unit[100]{\micron}$
 images. The colour scales are exaggerated to highlight the
 presence of the faint emission feature FIRS. The symbols mark the positions
 of SMM1 ($\medbullet$), IRS ($\bigstar$), SMM2 ($\blacksquare$), and
 the FHSC candidate MMS ($+$). The dashed line marks the FWHM of the SMM1 volume
 density distribution. In both images, the
 contours are plotted in steps of $3\sigma$ above the level
 of the background cloud. (This
 noise does not necessarily represent the true noise level
 around the protostar.) The beam size of the $\unit[100]{\micron}$ observations
 is indicated in the bottom right corner.}
\end{figure}

Among our available FIR data, the $\unit[100]{\micron}$ map offers the
highest spatial resolution, which is beneficial for disentangling the
emission of the Class~I source IRS and a potential emission
contribution from an embedded source in SMM1. In the original $\unit[100]{\micron}$
map, additional faint
emission close to the dominating signal from the Class I source is
apparent, but is confused by the non-trivial structure of the PACS
point-spread function (PSF). We therefore deconvolve the
$\unit[100]{\micron}$ map to obtain a
clearer view of IRS and its immediate surroundings.

In \citet{che12} (hereafter referred to as \citetalias{che12}),
we show a preliminary analysis where the \textit{Herschel}
$\unit[100]{\micron}$ data was reduced
using Scanamorphos v9.0 \citep{rou12} and the {\tt nothermal}, {\tt
 noglitch,} and {\tt galactic} options. For the deconvolution, which
was performed with the {\tt miriad} task {\tt CLEAN}, the azimuthally
averaged \textit{Herschel} PSF at $\unit[100]{\micron}$ \citep{ani11} was
taken. This treatment of the data
indicated a faint, slightly extended $\unit[100]{\micron}$ emission
source at a distance of $\sim\unit[20]{\arcsec}$ to the south-east of
IRS with a flux density of $\unit[36\pm2]{mJy}$ \citep[see
 Figure~2b and Table~4 of][]{che12}. The secondary
\textit{Herschel} source appeared to be spatially coincident
with the compact $\unit[1.3]{mm}$ continuum source, MMS. Its
SED -- including the $\unit[100]{\micron}$ source --
agreed with classifying it as a FHSC candidate embedded in
the prestellar core SMM1.

Here we reiterate 
the nature of this faint $\unit[100]{\micron}$ source -
hereafter called CB\,17-FIRS. We performed a more
thorough analysis, using the latest data reduction version
(Scanamorphos v18.0; {\tt nothermal} and {\tt galactic} options),
together with the appropriate $\unit[100]{\micron}$ PACS PSF
($\unit[20]{\arcsec\,s^{-1}}$ scan speed, not re-centred, no enhanced
drizzling), derived from observations of the asteroid Vesta and
provided by the PACS instrument team\footnote{{\tt
 ftp://ftp.sciops.esa.int/pub/hsc-calibration/
 PACS/PSF/PACSPSF\_PICC-ME-TN-033\_v2.0.tar.gz}}. Before the
deconvolution, the PSF was rotated counter-clockwise by
$\unit[260.45]{\deg}$ in order to match the orientation of the PSF present
in the CB\,17 PACS observations. We 
applied five iterations of a modified Richardson-Lucy deconvolution
\citep{ric72,luc74} as implemented in IDL.

In the resulting deconvolved map, we confirm the presence of a
secondary emission patch, which is clearly distinguishable from the
emission of the Class I source (Figure~\ref{f-05}). 
This improved analysis results
in a positional shift of FIRS by $\unit[7]{\arcsec}$ towards the west,
as compared to the preliminary analysis reported in \citetalias{che12}.
The reason for this is two-fold. First, the new data reduction shows a 
slightly different emission structure in the immediate vicinity
of IRS. The weak emission, which is clearly visible in the
original images (Figure~\ref{f-02}), is found further away
from the Class~I protostar in
the early data reduction. 
Second, the use of the
properly structured and oriented PSF instead of the azimuthally
averaged PSF approximation minimises residual emission features from the
bright source IRS, which can be mistaken as true emission features from the
faint source FIRS. This unfavourable morphology makes both the
data calibration and the image deconvolution challenging and
introduces uncertainties for position and
flux density beyond the level we would experience for an isolated source.

In contrast to its position, the determined flux density and size of
FIRS remained unchanged with respect\ to the preliminary
analysis. Owing to the proximity of IRS, the size and flux
density are estimated by fitting 1D Gaussian profiles along north-east
and south-west directions. We determine a FWHM of
$\unit[11.3]{\arcsec}$ (which is considerably larger than the beam
size of $\sim\unit[7.1]{\arcsec}$) and a flux density of
$\unit[35]{mJy}$. Considering the difficulties, which were discussed
in the previous paragraph, we assume that the flux density is not
known better than to a factor of 2. However,
according to our current analysis, we assume that we can
determine the position of FIRS fairly reliably. The implications of this newly derived position
on the FHSC candidate MMS is discussed in Section~\ref{s-dis_firs}.

\subsection{Kinematics of the dense core SMM1}

To analyse the kinematics of the dense core SMM1, we
obtained complementary observations of \nnh(1--0)
with an angular resolution of $\sim\unit[4]{\arcsec}$.
The hyperfine
structure complex offers the big advantage of providing a multitude of
parameters, such as optical depth, $\tau$, or excitation temperature,
$T_\mathrm{ex}$, with a single observation, for which usually
observations of optically thin and thick isotopes would be necessary.
Furthermore, the simultaneous fit of all HFS satellites allows
determination of the systemic velocity $v_\mathrm{LSR}$ and line FWHM
$\Delta v$ with much higher precision than for a single line, even for
data with moderate-to-low S/N. \citet{lip13} use single-dish observations
of \nnh{} in concert with CO data to determine molecular abundance profiles
and gas-phase depletion in CB\,17 and other starless cores of the EPoS-sample,
but they do not spatially resolve the kinematic structure as we attempt to do
here.

For the fitting, we used the hyperfine structure fitting routine
provided by {\tt CLASS}, which is a part of the software package
{\tt GILDAS}. The transition frequencies and relative weights of the
HFS lines were taken from \citet{cas95} and \citet{wom92},
respectively. For the fit we only included spectra, where the isolated
HFS component ($F_1F=01\rightarrow12$) exhibited a peak value of
$\geq5\sigma$.

\begin{figure*}[tb]
        \includegraphics{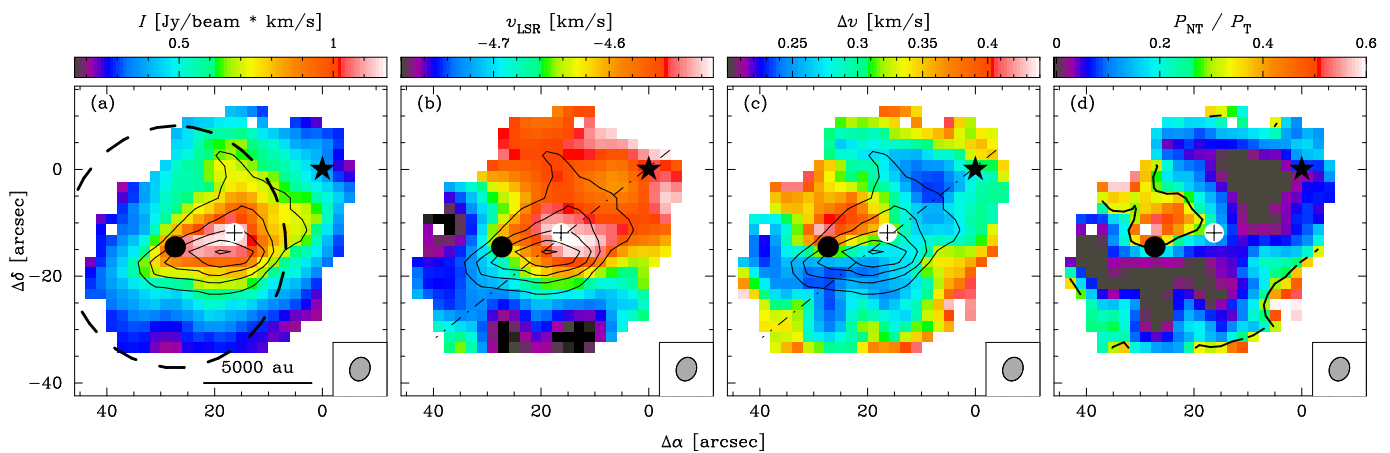}
        \caption{\label{f-06} (From left to right) \nnh{} maps
        of integrated intensity, velocity centroid, linewidth, and
        ratio of non-thermal to thermal pressure. Symbols mark the positions of SMM1
        ($\medbullet$), MMS ($+$), and IRS ($\bigstar$). The dashed
        circle in panel (a) represents the FWHM volume density distribution
        of SMM1. The contours in
        panels (a)-(c) represent the N$_2$H$^+$ column density with
        contours from $\unit[(1\ldots3)\times10^{13}]{cm^{-2}}$ in
        steps of $\unit[0.5\times10^{13}]{cm^{-2}}$. In panel (d) the
        contour marks the $P_\mathrm{NT}/P_\mathrm{T}=0.3$ line. The dash-dotted
        lines in panels (b) and (c) mark the cut for the
        position-velocity diagram, which is shown in
        Figure~\ref{f-07}. The synthesised beam size is indicated in the bottom right
        corners.}
\end{figure*}

The spectrum towards the column density peak of N$_2$H$^+$ exhibits a total
optical depth, i.e., the sum of the optical depths of the individual
HFS components, of $\tau=25$, an excitation temperature of
$T_\mathrm{ex}=\unit[7.5\pm0.3]{K}$, and a peak column density of
$N(\mathrm{N_2H^+})=\unit[3.6\times10^{13}]{cm^{-2}}$.
Maps of the
fitted \nnh{} integrated line intensity $I$, the velocity centroid
$v_\mathrm{LSR}$, linewidth $\Delta v$, and column density
$N(\mathrm{N_2H^+})$ are shown in
Figure~\ref{f-06}. Additionally, we plot the ratio of thermal
to non-thermal pressure \citep{taf04}
\begin{align}
        \frac{P_\mathrm{NT}}{P_\mathrm{T}}=\frac{\Delta v_\mathrm{NT}^2}{\sqrt{8\,\log(2)}}\,\frac{\sub{\mu}{p}\,m_\mathrm{H}}{k_\mathrm{B}\,T},
\end{align}
where $\sub{\mu}{p}=2.32$ is the mean atomic mass per particle in
molecular clouds \citep{prz08}. The
temperature was taken from the RT fit (Section~\ref{s-rt}),
because we assumed the gas and dust to be well-coupled in the centre at
densities $\sub{n}{H}\sim\unit[2\times10^5]{cm^{-3}}$. The
non-thermal linewidth was derived by
\begin{align}
        \Delta\sub{v}{NT}=\sqrt{\Delta v^2 - \left(\Delta\sub{v}{T}^2 + \Delta\sub{v}{instr}^2\right)},
\end{align}
where $\Delta v$ is the measured linewidth, $\Delta\sub{v}{T}$  the
thermal linewidth, and $\Delta\sub{v}{instr}=\unit[0.235]{km\,s^{-1}}$
is the instrumental resolution
(which is 1.875 times the channel spacing).

Clearly, \nnh(1--0) emission is only coming from within SMM1, whereas no
emission comes from IRS (Figure~\ref{f-06}a). The position
of the column density peak of \nnh{} can be associated with MMS,
which is offset from the centre of SMM1. This could be due to
substructure in the densest part, which can be traced
by interferometric molecular line observations, but remains undetected by the
large-scale dust emission at five times lower spatial resolution.
The \nnh{} column density distribution can be
approximated by an elliptical Gaussian with major and minor axes (FWHM)
of $\unit[7\,100]{au}$ and $\unit[5\,900]{au}$, respectively. The
major axis exhibits a $P.A.=\unit[135]{\deg}$ (East of
North).

We derive a N$_2$H$^+$ peak column density that is about ten times higher than the value determined by \citet{cas02a}.
Their low spatial resolution, however, leads to considerable beam
dilution, which then underestimates the excitation
temperature, optical depth, and thus the column density. From fitting the
IRAM\,30m single-dish spectra alone, we determine an optical depth of $\tau\sim9$
and $T_\mathrm{ex}\sim\unit[5]{K}$, which agrees with the
observations of \citet{cas02a}. To assess whether our interferometric
observations of N$_2$H$^+$ also suffer from beam
dilution effects, we estimated the expected excitation temperature
$\sub{T}{ex}$ with the on-line tool RADEX \citep{tak07}. We used the
best-fit dust temperature $T_0$, as well as the \nnh{} column density and
linewidth, as input parameters. The observed value of
$\sub{T}{ex}=\unit[7.5]{K}$ would be reached at a density of
$\sub{n}{H}\sim\unit[3\times10^5]{cm^{-3}}$, which is
in good agreement with our finding of the central density of
$\sub{n}{H,0}=\unit[2.3\times10^5]{cm^{-3}}$, taking
the uncertainties introduced by the dust model into account.
This analysis indicates that the $\unit[25]{\arcsec}$ resolution of
Herschel observations is sufficient to accurately sample the structure
of the core on the scales of the dense core plateau. This in
turn allows us to obtain a robust estimate of the central volume density.

The velocity map (Figure~\ref{f-06}b) shows a clear gradient
from south-east towards north-west of
$\sim\unit[4.3\pm0.2]{km\,s^{-1}\,pc^{-1}}$ \citep[following the
 recipe of][]{kan97}. In addition to the large-scale velocity
gradient, which we attribute to rotation of the core around an axis
with $P.A.=\unit[54\pm2]{\deg}$, the velocity map also exhibits a
complex small-scale structure, which is discussed in more detail in
Section~\ref{s-dis_velo}. 

Similar to the velocity profile, the linewidth map
(Figure~\ref{f-06}c) also reveals complex structure. Large
parts of the dense core exhibit a relatively constant line width of
the order of $\Delta v\sim\unit[0.25]{km\,s^{-1}}$.
Figure~\ref{f-06}d shows that non-thermal contributions
in some parts of the cloud are negligible. We furthermore see no evidence
of any infall motions, which would reveal themselves through increased linewidth
towards the centre \citep{cas02b} and -- in case of excitation
gradients along the line of sight in concert with optical depths above unity --
asymmetric, self-absorbed line profiles \citep{cra05}. This is consistent
with \citet{pav06}, who finds only weak infall motions in the envelope.
On the other hand, \nnh{} freeze-out towards the core centre could inhibit the
appearance of infall signatures due to the absence of gas phase molecules in
that region \citep{lip13}.

The only region with considerably elevated linewidths is a cone-shaped
region that opens up towards the north-east.
We find a ratio of non-thermal to thermal pressure
$P_\mathrm{NT}/P_\mathrm{T}>0.3$ (Figure~\ref{f-06}d) in this
region. We do not attribute the increased linewidth
to being caused by infall motions due to the lack of spherical symmetry.

\section{Discussion}
\label{s-discussion}

\subsection{A large-scale overview}
\label{s-dis-largescale}

In Section \ref{s-rt}, we determine the temperature
and column density distribution of CB\,17 via MBB and
RT fitting (Figure~\ref{f-03}). The cometary shape of
the globule stands out clearly. The column densities in the
\textit{cometary tail} towards the north-east are about
ten times lower than the peak column density of the core.
Interestingly, towards the
north-west we see a warm \textit{west wing}, where the dust temperatures
are $\gtrsim\unit[20]{K}$ and hydrogen column densities are as low
as $\sim\unit[10^{19}]{cm^{-2}}$. 
As discussed earlier,
the specific morphology of CB17 and the low envelope 
temperature (as compared to most other globules) prevent us from deriving 
reliable values for the extended envelope levels of both the temperature and 
column density.

In the \textit{cometary tail} we already reach our sensitivity limits at
column densities of the order of
$\sub{N}{H}\sim\unit[3\times10^{20}]{cm^{-2}}$, hence more
than a magnitude higher than the \textit{west wing}. This is due
to the lower temperatures and thus the lower emission levels. The
\textit{west wing} might be part of a warm background/foreground cloud or filament
that is seen in the large-scale $\unit[160]{\micron}$ map in
Figure~\ref{f-02}. Whether this region is physically connected
to CB\,17 is not clear. An answer to this question could be provided by
molecular line data that
traces the kinematics across the boundary, which we do not have at hand.

\subsection{Characterising SMM1}
\label{s-dis_cb17smm1a}

\subsubsection{Energy balance}

Starless and prestellar cores are the best probes for a ray-tracing fitting
analysis as introduced in Section~\ref{s-rt} thanks to their relatively
simple temperature and volume density structure. 
Knowing the temperature and density profiles allows us to perform a
stability analysis of the dense core SMM1 at an unprecedented level of
accuracy. Using the equations listed in Appendix~\ref{s-appendix},
we find a gravitational potential energy of
$\sub{E}{grav}=-\unit[(4.2\pm1.0)\times10^{35}]{J}$, a thermal energy
of $\sub{E}{therm}=\unit[(3.0\pm0.5)\times10^{35}]{J}$, a turbulent
energy of $\sub{E}{turb}=\unit[(2.3\pm0.4)\times10^{34}]{J}$ (where
we estimated an average non-thermal linewidth of
$\Delta\sub{v}{NT}=\unit[0.2]{km\,s^{-1}}$), and a rotational energy
of $\sub{E}{rot}=\unit[(6.3\pm1.8)\times10^{34}]{J}$ within the central
$\unit[18,000]{au}$.

Using the results from above we also assess the stability of
SMM1 by comparing the various energy components. Thanks to the availability
of kinematic information, we are not limited to comparing gravitational
potential and thermal energies \citep[e.g.][]{lip13}. By neglecting the
contribution from magnetic fields
for the moment, the criterion for a core to be bound is
\begin{align}
        \label{e-04}
        \frac{\sub{E}{therm}+\sub{E}{turb}+\sub{E}{rot}}{|\sub{E}{grav}|}\leq1.
\end{align}

Integration of the density and temperature
distribution from inside out shows that this ratio is $\gg1$ at small
radii, reaches unity at $r\sim\unit[8\,000]{au}$, and
converges towards $0.93\pm0.10$ at $\sub{r}{core}$. This means
that the core is literally on the margin of being bound at the scale of
the core size.
\citet{pav06} find a chemical age of the
envelope of CB\,17 of $\unit[2]{Myr}$, which is five times its free-fall
time. Therefore, these authors conclude that CB\,17 must have evolved
quasi-statically for some time, which agrees with our result
of a globule on the verge of being bound -- neither collapsing nor
dispersing.

This picture will also not change when including magnetic fields.
An attempt to measure
the degree of polarisation, hence the magnetic field strength in CB\,17,
has been undertaken by \citet{mat09}. Although we are not able to get a statistically
significant estimate of the magnetic field
with only two polarisation vectors, we can use these observations
to estimate an upper limit of the order of $\unit[10]{\mu G}$, and
consequently a magnetic energy of the order of
$\sub{E}{mag}\lesssim\unit[3\times10^{34}]{J}$.
Therefore, the influence of magnetic fields on the stability
of CB\,17 is considered to be insignificant.

Using the LoS-averaged column density and temperature
maps results in over-predicting the thermal energy by a factor of 1.3.
In addition, the LoS-averaged column density map does not yield a direct
handle on the value of the gravitational potential
$\sub{E}{grav}=\sub{\alpha}{vir}\,G\sub{M}{core}^2/\sub{r}{core}$,
since the pre-factor $\sub{\alpha}{vir}$ depends on the radial volume density
distribution of mass. Thus, the ray-tracing allows us to significantly
reduce the uncertainties in our parameter estimates, greatly
improving the ability to assess the energy balance of SMM1.
Moreover, including the information about kinematics (turbulence plus
rotation) allows us to get a more reliable estimate of the core state, since
its contribution is about 30\% of the thermal energy. Simply neglecting it would
lead to the conclusion that SMM1 is indeed bound.

\begin{figure}[tb]
 \includegraphics{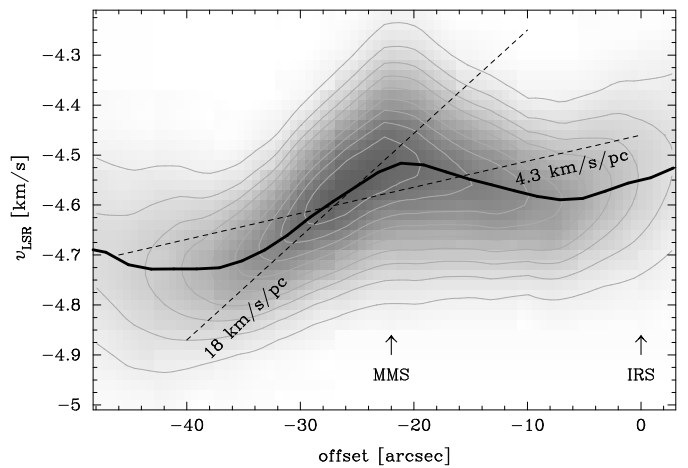}
 \caption{\label{f-07} Position-velocity diagram for
  \nnh(1--0) along the cut marked in
  Figure~\ref{f-06}b,c from south-east to
  north-west. The solid line marks the fitted velocity to
  highlight the shape of the curve. Contours are plotted in steps of
 10\% of the peak emission. The two dashed lines indicate (a) the
 velocity gradient attributed to solid body rotation of the core
 (with a measured slope of $\unit[4.3]{km\,s^{-1}\,pc^{-1}}$), and (b) the
 CAD12 N$_2$D$^+$(3-2) gradient of $\unit[18]{km\,s^{-1}\,pc^{-1}}$ for comparison.
 See Section~\ref{s-dis_velo} for more details.}
\end{figure}

\subsubsection{Velocity structure}
\label{s-dis_velo}

Inspection of the velocity map (Figure~\ref{f-06}b) already
reveals a high level of complexity. Fitting the large-scale velocity distribution
reveals a gradient of $\unit[4.3]{km\,s^{-1}\,pc^{-1}}$
from south-east to north-west. This is superposed by a sinusoidal pattern with a
peak-to-peak amplitude of $\sim\unit[0.15]{km\,s^{-1}}$
(Figure~\ref{f-07}). We interpret this velocity structure as being
the result of the superposition of two types of motion: SMM1
exhibits solid-body rotation on a large scale, but the inner regions
are characterised by differential motions with respect to the dense core.
In this scenario, the blue-shifted material
with respect to the dense \nnh{} knot represents the outer layers
that get stripped off
due to the differential motion, which is $\sim\unit[0.2]{km\,s^{-1}}$.
In their survey of 42~isolated cores, \citet{wal04} find 
only small velocity differences $\lesssim\unit[0.1]{km\,s^{-1}}$
between \nnh{} cores and envelopes
(traced by $^{13}$CO and C$^{18}$O). For CB\,17, these authors find a velocity
difference of $\sim\unit[0.1]{km\,s^{-1}}$ between \nnh{} and C$^{18}$O,
whereas $^{13}$CO and \nnh{} are
at about the same velocity. However, their low spatial resolution
of $\sim\unit[50]{\arcsec}$ does not allow them to resolve the small-scale
structure of this source. An additional complication comes from the
fact that CB\,17 has fragmented into at least two sources.
It could be that the motion of the \nnh{} knot
within the dense core is part of an orbital
motion with IRS. We have no means to quantify
that assumption, but from the
observations of molecular lines at hand, we can assume an upper
limit of the relative radial
velocities of $\unit[0.2]{km\,s^{-1}}$, which is a value that does not
contradict what can be expected for the orbital motion of two
bodies with a few solar masses at a separation of the order of
$\unit[10^4]{au}$.

The observed overall velocity gradient of $\unit[4.3]{km\,s^{-1}\,pc^{-1}}$
 agrees with what is
generally found for prestellar cores \citep{cas02a}, whereas Class~0 sources
rotate an order of magnitude faster \citep{che07}. In their observations of
N$_2$D$^+$(3--2), \citetalias{che12} derived a velocity gradient of
$\unit[18]{km\,s^{-1}\,pc^{-1}}$, which led to the conclusion that SMM1 should
already be a rather evolved core and might have formed a FHSC, MMS,
in the centre. However, N$_2$D$^+$(3--2) is only observed
(due to missing short spacings in the interferometric observations,
molecular abundance, and/or excitation conditions) in a region
to the south-east of the N$_2$H$^+$ column density peak. This region
translates into offsets between $-35$ and $-\unit[20]{\arcsec}$ in the
position-velocity cut (Figure~\ref{f-07}), where the velocity profile
of N$_2$H$^+$ is also rather steep. Therefore, we conclude that
N$_2$D$^+$ only traces parts of the core velocity profile.
Following the velocity structure throughout
the whole core reveals an overall rotation structure,
which is somewhat characteristic of prestellar cores.

The velocity structure is, however, not only characterised by a complex
rotation pattern, but the observations of N$_2$H$^+$ also reveal a surprising
pattern in the linewidth. In large parts of the core, the linewidth can be
fully explained with only thermal motions, but a region towards the north-east of
SMM1 shows a relatively high non-thermal contribution. This could be induced
through outflow activity from the nearby protostar IRS, but also through
a possible low-velocity outflow from MMS (\citetalias{che12}).

\subsection{SMM1 -- starless or protostellar?}
\label{s-dis_firs}

In the previous sections we have established that SMM1 is a slowly
rotating dense core in a rather young evolutionary stage (characterised by
a flattening of the volume density profile towards the core centre)
that is on the verge of being bound. However, the question of whether
SMM1 has a starless or a protostellar nature remains.

The key to this question can be the $\unit[100]{\mu m}$ feature FIRS. Its
position and flux density from a preliminary analysis fitted
well into the picture of a FHSC at the position of MMS (\citetalias{che12}).
Using our latest calibration and data reduction techniques, we derive a
new, more reliable position of FIRS. This new analysis suggests that FIRS
is actually offset from MMS by
$\sim\unit[7]{\arcsec}$. Thanks to the clear detection of IRS in both the
$\unit[100]{\mu m}$ and the $\unit[1.3]{mm}$ continuum image of
\citetalias{che12}, we can align these two maps to sub-arcsec precision.
Additionally, the clear detection of FIRS
($\gg\unit[3]{\sigma}$, Figure~\ref{f-05}) allows
us to reach a positional accuracy down to
a fraction of the beam size. This suggests that MMS and FIRS are two unrelated
features.

This immediately leads to the
question of the origin of FIRS. One option is that FIRS is instead
associated with both the NIR nebula to the south-east of IRS, and the north-western
extension of the large-scale mm-continuum emission SMM2
(see Figure~\ref{f-01}). These three observational features
could originate in the working surface associated with the outflow from IRS. In
this scenario, the blue-shifted outflow velocities would imply
that IRS resides on the far side of the
CB\,17 cloud. The low-velocity dispersion of N$_2$H$^+$ at the position of
FIRS would mean that the working surface must be located outside the
\nnh{} emission region at the core edge.
Further along its path, the deflected outflow then could penetrate the dense core
and stir up material, providing the non-thermal line broadening to the north-east of
SMM1.

Thus, the marginal SMA detection of MMS remains the only indication that the
dense core SMM is possibly no longer starless in nature, but already
contains an embedded FHSC. That MMS can be associated with the
position of the \nnh{} column density peak
contradicts the FHSC hypothesis at first glance.
The warming up of the central regions should result in the
release of CO from the icy grain mantles, which would lead to the destruction
of \nnh{}. However, if the FHSC has only formed recently, it does not necessarily
have to be reflected in an immediately observable change in the \nnh{} structure.
While the new analysis of the Herschel data
presented here could not provide additional supporting evidence, it also does
not contradict the still unconfirmed FHSC hypothesis. Therefore,
the question of whether
SMM1 is starless or protostellar remains unanswered until new observational
evidence can be presented.

\section{Summary}
\label{s-summary}

In our study of the isolated Bok globule CB\,17, we used FIR and
(sub-)mm thermal emission to determine the dust temperature and
density structure of CB\,17. Interferometric observations of the
\nnh(1--0) transition allowed us to determine the kinematic state of
the dense core SMM.

The main conclusions are as follows:
\begin{itemize}
\item Using LoS-averaged and ray-tracing fitting routines, we determine
the column density and dust temperature distribution of the
cometary globule CB\,17. We found a peak column density of
$\sub{N}{H}=\unit[(4.3\pm0.3)\times10^{22}]{cm^{-2}}$. The core is
approximately spherically symmetric (correcting for
ellipticity and excluding IRS from that view) in
the central $\sim\unit[18\,000]{au}$, but we
observed a progressive decrease in symmetry as
the radius increases. We distinguished three regions: The
\textit{core} in the centre, the hot \textit{west wing}, and the \textit{cometary tail}.
\item For the \textit{core} region (SMM1), we derived radial
temperature and volume density profiles using a RT
fitting procedure. We found a mass of $\unit[2.3\pm0.3]{\msun}$.
The temperature is found to drop to
$\sim\unit[10.6\pm0.3]{K}$ in the centre, where we found a
central density of
$\sub{n}{H,0}\sim\unit[(2.3\pm0.2)\times10^5]{cm^{-3}}$.
The density profile exhibits a flat density plateau out to
$r_0=\unit[(9.5\pm0.6)\times10^3]{au}$. An analysis of the temperature and
density profiles shows
that on a global scale SMM1 is at the margin of being
gravitationally bound. This could be the reason it is
-- although chemically evolved -- dynamically still in a very early
evolutionary stage.
\item The rotation curve of the dense core cannot be explained
in simple terms, but we interpreted it as a complex superposition of
solid body rotation, together with internal motions of the densest regions.
\item We substantiated the finding of FIRS, a small,
slightly extended $\unit[100]{\micron}$
emission feature at a projected separation of
$\sim\unit[3\,400]{au}$ ($\sim\unit[13.5]{\arcsec}$) from IRS. Its
position differs slightly from our preliminary result reported
in \citetalias{che12}, which is caused by using the
appropriate \textit{Herschel} PSF and a more recent data reduction
software. Overall, we found weakened evidence for a FHSC in CB\,17. We
argued that the sources MMS (\citetalias{che12}) and FIRS are unrelated, but
we cannot determine the true nature of either of these two sources due to the
overall complexity of this source and the proximity to IRS.
\end{itemize}

Looking at the evidence provided
by the observations presented in this paper, we can conclude that CB\,17 is
a source that is in a
very early stage of star formation. It remains a key source
for studying the conditions of star formation just prior to or in a very
early stage of the protostellar collapse. Future
observations with upcoming facilities like NOEMA will help shed more
light onto the puzzle of CB\,17.

\begin{acknowledgements}
MS acknowledges support from NOVA, the Netherlands Research School for Astronomy.
This work was supported by NSF grants 0708158 and 0953142.
The work of AMS was supported by the Deutsche Forschungsgemeinschaft
priority programme 1573 ("Physics of the Interstellar Medium").
HL was funded by the Deutsches Zentrum für Luft- und Raumfahrt (DLR).
This research
used the facilities of the Canadian Astronomy Data Centre operated by
the National Research Council of Canada with the support of the
Canadian Space Agency. The James Clerk Maxwell Telescope is operated
by the Joint Astronomy Centre on behalf of the Science and Technology
Facilities Council of the United Kingdom, the Netherlands Organisation
for Scientific Research, and the National Research Council of Canada. MS
acknowledges the use of the GILDAS software package
({\tt http://www.iram.fr/IRAMFR/GILDAS/}).
\end{acknowledgements}

\appendix

\section{Energy terms}
\label{s-appendix}

\subsection*{Gravitational potential energy}

The gravitational potential energy is given by
\begin{align}
 \sub{E}{grav}=-4\pi\,G\,\int_0^{\sub{r}{core}}\,\sub{M}{r}(r)\,\sub{n}{H}(r)\,\sub{\mu}{H}\,\sub{m}{H}\,r\,\dd{r},
\end{align}
where $G$ is the gravitational constant, $\sub{\mu}{H}=1.40$ is the mean mass
per hydrogen atom \citep{prz08}, $\sub{m}{H}$ is the proton mass, and
\begin{align}
 \sub{M}{r}(r)=4\pi\,\sub{\mu}{H}\,\sub{m}{H}\,\int_0^r r^2 \sub{n}{H}(r)\,\dd{r}.
\end{align}
is the mass enclosed by a shell with radius $r$.

The gravitational potential energy is sometimes expressed as
$\sub{E}{grav}=-\sub{\alpha}{vir}\,G\,\sub{M}{core}^2/\sub{r}{core}$. Using
the volume density profiles derived in the ray-tracing fit, we
determine the pre-factor for CB\,17 to be $\sub{\alpha}{vir}=0.82\pm0.05$.

\subsection*{Thermal energy}
The thermal energy is given by
\begin{align}
 \sub{E}{therm}=\frac{3\,\sub{k}{B}}{2}\,4\pi\,\int_0^{\sub{r}{core}}\,\sub{n}{p}(r)\,T(r)\,r^2\,\dd{r},
\end{align}
where $\sub{k}{B}$ is the Boltzmann constant, and
$\sub{n}{p}=\sub{n}{H}\,\sub{\mu}{H}/\sub{\mu}{p}$ is the volume density of
particles with $\sub{\mu}{p}=2.32$ being the mean atomic mass per
particle in molecular clouds \citep{prz08}.

\subsection*{Rotational energy}
The energy for solid-body rotation at an angular frequency $\omega$
is
\begin{align}
 \sub{E}{rot}=\frac{1}{2}\,I\omega^2.
\end{align}
The moment of inertia $I$ is can be calculated by
\begin{align}
 I &= \int_V \sub{\mu}{H}\,\sub{m}{H}\,\sub{n}{H}(r)\,s^2\,\dd{V}
\end{align}
with $s$  the distance to the rotation axis. Assuming a spherical coordinate system and rotation around the axis
with polar angle $\vartheta=0$ we obtain $s=r\,\sin\vartheta$. Hence, the moment of inertia is
\begin{align}
 I &= 2\pi\,\sub{\mu}{H}\,\sub{m}{H}\,\int_0^{\sub{r}{core}} \int_0^\pi\, r^4\,\sin^3\vartheta\,\sub{n}{H}(r)\,\dd{\vartheta}\dd{r}\\
 &= \frac{8\pi}{3}\,\sub{\mu}{H}\,\sub{m}{H}\,\int_0^{\sub{r}{core}}\,r^4\,\sub{n}{H}(r)\,\dd{r}.
\end{align}

\subsection*{Turbulent energy}
Assuming a Boltzmann-distributed turbulent velocity, the turbulent energy content
can be estimated by relating the squared non-thermal line broadening
$\Delta \sub{v}{NT}^2$ to the averaged squared 3D-velocity
$\langle v^2 \rangle$. The kinetic energy of the turbulent gas is then
given by
\begin{align}
 \sub{E}{turb} &= \frac{\sub{M}{core}\,\langle v^2\rangle}{2}\\[2ex]
 &=\frac{3\,\sub{M}{core}\,\Delta \sub{v}{NT}^2}{16\,\ln(2)}.
\end{align}

\subsection*{Magnetic energy}

Assuming a constant
magnetic field strength throughout the core, the magnetic energy is
\begin{align}
 \sub{E}{mag}&=\frac{4\pi\,\sub{r}{core}^3}{3}\,\sub{u}{mag},
\end{align}
where $\sub{u}{mag}=B^2/(2\,\mu_0)$ is the magnetic
energy density, $B$  the magnetic field, and $\mu_0$ the magnetic
permeability constant.


\begin{thebibliography}{}
\bibitem[Andr{\'e} et al.(2010)]{and10} Andr{\'e}, P., et al.\ 2010, \aap, 518, L102 
\bibitem[Andr{\'e} et al.(2013)]{and13} Andr{\'e}, P., Di Francesco, J., Ward-Thompson, D., et al.\ 2013, arXiv:1312.6232 
\bibitem[Aniano et al.(2011)]{ani11} Aniano, G., Draine, B.~T., Gordon, K.~D., \& Sandstrom, K.\ 2011, \pasp, 123, 1218
\bibitem[Arzoumanian et al.(2011)]{arz11} Arzoumanian, D., et al.\ 2011, \aap, 529, L6
\bibitem[Belloche et al.(2006)]{bel06} Belloche, A., Parise, B., van der Tak, F.~F.~S., Schilke, P., Leurini, S., G{\"u}sten, R., \& Nyman, L.-{\AA}.\ 2006, \aap, 454, L51
\bibitem[Beuther et al.(2010)]{beu10} Beuther, H., Henning, T., Linz, H., Krause, O., Nielbock, M., \& Steinacker, J.\ 2010, \aap, 518, L78 
\bibitem[Beuther et al.(2013)]{beu13} Beuther, H., et al.\ 2013, \aap, 553, A115
\bibitem[Bergin \& Tafalla(2007)]{ber07} Bergin, E.~A., \& Tafalla, M.\ 2007, \araa, 45, 339
\bibitem[Caselli et al.(1995)]{cas95} Caselli, P., Myers, P.~C., \& Thaddeus, P.\ 1995, \apjl, 455, L77
\bibitem[Caselli et al.(2002a)]{cas02a} Caselli, P., Benson, P.~J., Myers, P.~C., \& Tafalla, M.\ 2002a, \apj, 572, 238
\bibitem[Caselli et al.(2002b)]{cas02b} Caselli, P., Walmsley, C.~M., Zucconi, A., Tafalla, M., Dore, L., \& Myers, P.~C.\ 2002b, \apj, 565, 331
\bibitem[Caselli et al.(2012)]{cas12} Caselli, P., et al.\ 2012, \apjl, 759, L37
\bibitem[Chen et al.(2007)]{che07} Chen, X., Launhardt, R., \& Henning, T.\ 2007, \apj, 669, 1058
\bibitem[Chen et al.(2010)]{che10} Chen, X., Arce, H.~G., Zhang, Q., Bourke, T.~L., Launhardt, R., Schmalzl, M., \& Henning, T.\ 2010, \apj, 715, 1344
\bibitem[Chen et al.(2012)]{che12} Chen, X., Arce, H.~G., Dunham, M.~M., Zhang, Q., Bourke, T.~L., Launhardt, R., Schmalzl, M., \& Henning, T.\ 2012, \apj, 751, 89
\bibitem[Clemens \& Barvainis(1988)]{cle88} Clemens, D.~P., \& Barvainis, R.\ 1988, \apjs, 68, 257 
\bibitem[Commer{\c c}on et al.(2012)]{com12} Commer{\c c}on, B., Launhardt, R., Dullemond, C., \& Henning, T.\ 2012, \aap, 545, A98
\bibitem[Crapsi et al.(2005)]{cra05} Crapsi, A., Caselli, P., Walmsley, C.~M., Myers, P.~C., Tafalla, M., Lee, C.~W., \& Bourke, T.~L.\ 2005, \apj, 619, 379
\bibitem[Enoch et al.(2010)]{eno10} Enoch, M.~L., Lee, J.-E., Harvey, P., Dunham, M.~M., \& Schnee, S.\ 2010, \apjl, 722, L33
\bibitem[Evans et al.(2001)]{eva01} Evans, N.~J., II, Rawlings, J.~M.~C., Shirley, Y.~L., \& Mundy, L.~G.\ 2001, \apj, 557, 193
\bibitem[Evans et al.(2009)]{eva09} Evans, N.~J., II, et al.\ 2009, \apjs, 181, 321
\bibitem[di Francesco et al.(2007)]{dif07} di Francesco, J., Evans, N.~J., II, Caselli, P., Myers, P.~C., Shirley, Y., Aikawa, Y., \& Tafalla, M.\ 2007, Protostars and Planets V, 17
\bibitem[Griffin et al.(2010)]{gri10} Griffin, M.~J., et al.\ 2010, \aap, 518, L3 
\bibitem[Henning et al.(2010)]{hen10} Henning, T., Linz, H., Krause, O., Ragan, S., Beuther, H., Launhardt, R., Nielbock, M., \& Vasyunina, T.\ 2010, \aap,518, L95 
\bibitem[Johnstone et al.(2010)]{joh10} Johnstone, D., Rosolowsky, E., Tafalla, M., \& Kirk, H.\ 2010, \apj, 711, 655
\bibitem[J{\o}rgensen et al.(2002)]{jor02} J{\o}rgensen, J.~K., Sch{\"o}ier, F.~L., \& van Dishoeck, E.~F.\ 2002, \aap, 389, 908
\bibitem[Kane \& Clemens(1997)]{kan97} Kane, B.~D., \& Clemens, D.~P.\ 1997, \aj, 113, 1799
\bibitem[Kainulainen et al.(2013)]{kai13} Kainulainen, J., Ragan, S.~E., Henning, T., \& Stutz, A.\ 2013, arXiv:1305.6383 
\bibitem[Lada \& Lada(2003)]{lad03} Lada, C.~J., \& Lada, E.~A.\ 2003, \araa, 41, 57
\bibitem[Launhardt \& Henning(1997)]{lau97} Launhardt, R., \& Henning, T.\ 1997, \aap, 326, 329
\bibitem[Launhardt et al.(2010)]{lau10} Launhardt, R., et al.\ 2010, \apjs, 188, 139
\bibitem[Launhardt et al.(2013)]{lau13} Launhardt, R., et al.\ 2013, \aap, 551, A98
\bibitem[Larson(1969)]{lar69} Larson, R.~B.\ 1969, \mnras, 145, 271
\bibitem[Lemme et al.(1996)]{lem96} Lemme, C., Wilson, T.~L., Tieftrunk, A.~R., \& Henkel, C.\ 1996, \aap, 312, 585
\bibitem[Leung(1975)]{leu75} Leung, C.~M.\ 1975, \apj, 199, 340
\bibitem[Linz et al.(2010)]{lin10} Linz, H., et al.\ 2010, \aap, 518, L123
\bibitem[Lippok et al.(2013)]{lip13} Lippok, N., et al.\ 2013, \aap, 560, A41
\bibitem[Lucy(1974)]{luc74} Lucy, L.~B.\ 1974, \aj, 79, 745
\bibitem[Lynds(1962)]{lyn62} Lynds, B.~T.\ 1962, \apjs, 7, 1 
\bibitem[Masunaga et al.(1998)]{mas08} Masunaga, H., Miyama, S.~M., \& Inutsuka, S.-I.\ 1998, \apj, 495, 346
\bibitem[Matthews et al.(2009)]{mat09} Matthews, B.~C., McPhee, C.~A., Fissel, L.~M., \& Curran, R.~L.\ 2009, \apjs, 182, 143
\bibitem[Molinari et al.(2010)]{mol10} Molinari, S., et al.\ 2010, \aap, 518, L100 2007, \apjl, 662, L23
\bibitem[Motte et al.(1998)]{mot98} Motte, F., Andre, P., \& Neri, R.\ 1998, \aap, 336, 150
\bibitem[Nielbock et al.(2012)]{nie12} Nielbock, M., et al.\ 2012, \aap, 547, A11
\bibitem[Ossenkopf \& Henning(1994)]{oss94} Ossenkopf, V., \& Henning, T.\ 1994, \aap, 291, 943 
\bibitem[Omukai(2007)]{omu07} Omukai, K.\ 2007, \pasj, 59, 589
\bibitem[Pavlyuchenkov et al.(2006)]{pav06} Pavlyuchenkov, Y., Wiebe, D., Launhardt, R., \& Henning, T.\ 2006, \apj, 645, 1212
\bibitem[Pavlyuchenkov et al.(2007)]{pav07} Pavlyuchenkov, Y., Henning, T., \& Wiebe, D.\ 2007, \apjl, 669, L101
\bibitem[Pezzuto et al.(2012)]{pez12} Pezzuto, S., et al.\ 2012, \aap, 547, A54
\bibitem[Pilbratt et al.(2010)]{pil10} Pilbratt, G.~L., et al.\ 2010, \aap, 518, L1 
\bibitem[Pineda et al.(2011)]{pin11} Pineda, J.~E., et al.\ 2011, \apj, 743, 201
\bibitem[Pitann et al.(2013)]{pit13} Pitann, J., et al.\ 2013, \apj, 766, 68
\bibitem[Poglitsch et al.(2010)]{pog10} Poglitsch, A., et al.\ 2010, \aap, 518, L2
\bibitem[Przybilla et al.(2008)]{prz08} Przybilla, N., Nieva, M.-F., \& Butler, K.\ 2008, \apjl, 688, L103
\bibitem[Ragan et al.(2012)]{rag12} Ragan, S., et al.\ 2012, \aap, 547, A49
\bibitem[Richardson(1972)]{ric72} Richardson, W.~H.\ 1972, Journal of the Optical Society of America (1917-1983), 62, 55
\bibitem[Roussel(2012)]{rou12} Roussel, H.\ 2012, arXiv:1205.2576 
\bibitem[Saigo et al.(2008)]{sai08} Saigo, K., Tomisaka, K., \& Matsumoto, T.\ 2008, \apj, 674, 997
\bibitem[Shirley et al.(2000)]{shi00} Shirley, Y.~L., Evans, N.~J., II, Rawlings, J.~M.~C., \& Gregersen, E.~M.\ 2000, \apjs, 131, 249
\bibitem[Shu(1977)]{shu77} Shu, F.~H.\ 1977, \apj, 214, 488
\bibitem[Sodroski et al.(1997)]{sod97} Sodroski, T.~J., Odegard, N., Arendt, R.~G., Dwek, E., Weiland, J.~L., Hauser, M.~G., \& Kelsall, T.\ 1997, \apj, 480, 173
\bibitem[Stutz et al.(2010)]{stu10} Stutz, A., et al.\ 2010, \aap, 518, L87
\bibitem[Stutz et al.(2013)]{stu13} Stutz, A.~M., et al.\ 2013, \apj, 767, 36
\bibitem[van der Tak et al.(2007)]{tak07} van der Tak, F.~F.~S., Black, J.~H., Sch{\"o}ier, F.~L., Jansen, D.~J., \& van Dishoeck, E.~F.\ 2007, \aap, 468, 627
\bibitem[Tafalla et al.(2004)]{taf04} Tafalla, M., Myers, P.~C., Caselli, P., \& Walmsley, C.~M.\ 2004, \aap, 416, 191
\bibitem[Taquet et al.(2013)]{taq13} Taquet, V., Peters, P.~S., Kahane, C., et al.\ 2013, \aap, 550, A127 
\bibitem[Tomida et al.(2010)]{tom10} Tomida, K., Tomisaka, K., Matsumoto, T., Ohsuga, K., Machida, M.~N., \& Saigo, K.\ 2010, \apjl, 714, L58
\bibitem[Walsh et al.(2004)]{wal04} Walsh, A.~J., Myers, P.~C., \& Burton, M.~G.\ 2004, \apj, 614, 194
\bibitem[Ward-Thompson et al.(1994)]{war94} Ward-Thompson, D., Scott, P.~F., Hills, R.~E., \& Andre, P.\ 1994, \mnras, 268, 276
\bibitem[Ward-Thompson et al.(1999)]{war99} Ward-Thompson, D., Motte, F., \& Andre, P.\ 1999, \mnras, 305, 143
\bibitem[Ward-Thompson et al.(2002)]{war02} Ward-Thompson, D., Andr{\'e}, P., \& Kirk, J.~M.\ 2002, \mnras, 329, 257
\bibitem[Womack et al.(1992)]{wom92} Womack, M., Ziurys,\ L.~M., \& Wyckoff, S.\ 1992, \apj, 387, 417
\bibitem[Young et al.(2004)]{you04} Young, C.~H., et al.\ 2004, \apjs, 154, 396
\bibitem[Zucconi et al.(2001)]{zuc01} Zucconi, A., Walmsley, C.~M., Galli, D.\ 2001, \aap, 376, 650
\end{thebibliography}
\end{document}